\documentclass[useAMS,usenatbib]{mn2e}

\usepackage{graphicx}

\newcommand\rs[1]{_\mathrm{#1}}
\def\gsim{\;\lower4pt\hbox{${\buildrel\displaystyle >\over\sim}$}\,}

\title[Accretion to young stars triggered by flares]{Mass accretion
to young stars triggered by flaring activity in circumstellar
disks}

\author[S. Orlando et al.]{Salvatore Orlando$^{1}$
\thanks{E-mail: orlando@astropa.inaf.it},
 Fabio Reale$^{2,1}$, Giovanni Peres$^{2,1}$, Andrea Mignone$^{3}$\\ \\
$^{1}$INAF - Osservatorio Astronomico di Palermo,
       Piazza del Parlamento 1, 90134, Palermo, Italy\\
$^{2}$Dip. di Fisica, Universit\`a degli
       Studi di Palermo, Piazza del Parlamento 1, 90134, Palermo,
       Italy\\
$^{3}$Dip. di Fisica Generale, Universit\`a degli Studi di
       Torino, via Pietro Giuria 1, 10125, Torino, Italy
}
\begin{document}

\date{Accepted. Received}

\pagerange{\pageref{firstpage}--\pageref{lastpage}} \pubyear{}

\maketitle

\label{firstpage}

\begin{abstract}
Young low-mass stars are characterized by ejection of collimated
outflows and by circumstellar disks which they interact with through
accretion of mass. The accretion builds up the star to its final mass
and is also believed to power the mass outflows, which may in turn
remove the excess angular momentum from the star-disk system. However,
although the process of mass accretion is a critical aspect of star
formation, some of its mechanisms are still to be fully understood. A
point not considered to date and relevant for the accretion process is
the evidence of very energetic and frequent flaring events in these
stars. Flares may easily perturb the stability of the disks, thus
influencing the transport of mass and angular momentum. Here we report
on three-dimensional magnetohydrodynamic modeling of the evolution of a
flare with an idealized non--equilibrium initial condition occurring near
the disk around a rotating magnetized star. The model takes into account
the stellar magnetic field, the gravitational force, the viscosity of
the disk, the magnetic-field-oriented thermal conduction (including the
effects of heat flux saturation), the radiative losses from optically
thin plasma, and the coronal heating. We show that, during its first
stage of evolution, the flare gives rise to a hot magnetic loop linking
the disk to the star. The disk is strongly perturbed by the flare: disk
material evaporates under the effect of the thermal conduction and an
overpressure wave propagates through the disk. When the overpressure
reaches the opposite side of the disk, a funnel flow starts to develop
there, accreting substantial disk material onto the young star from the
side of the disk opposite to the flare.
\end{abstract}


\section{Introduction}

Classical T Tauri Stars (CTTSs) are young low-mass stars actively
accreting mass from a surrounding disk (\citealt{1998apsf.book.....H,
2007IAUS..243.....B}). On the basis of the largely accepted magnetospheric
accretion scenario (\citealt{Koenigl1991ApJ}), the accretion process from
the disk is regulated by the stellar magnetic field which is strong enough
to disrupt the inner part of the disk at a distance of a few stellar radii
(the truncation radius) where the magnetic pressure is approximately equal
to the total gas pressure. The field guides the circumstellar material
along its flux tubes toward the central protostar, around free-fall
velocity, terminating in a shock at the photosphere. This scenario is
supported by much observational evidence, among others, the photospheric
magnetic field of a few kG that has been detected in a number of CTTSs
by exploiting the Zeeman effect (\citealt{1999ApJ...510L..41J}).

One of the fundamental issues in the magnetospheric accretion
scenario is the mechanism of inward transport of matter and outward
transport of angular momentum needed to explain the final mass and
angular momentum of mature stars. The turbulence in the disk has
been proposed to be the main responsible of the required outward
angular momentum transport, thus controlling the mass accretion on the
central star (\citealt{1973A&A....24..337S}). Nowadays, there is a large
consensus in the literature that the turbulence is mainly driven by
the magnetorotational instability (MRI; \citealt{1991ApJ...376..214B,
1998RvMP...70....1B}). MRI operates in regions where the ionization
is sufficient to couple the gas to the magnetic field and leads to the
transfer of angular momentum along the magnetic field lines connecting gas
located in different orbits (\citealt{1995ApJ...440..742H, 1995ApJ...446..741B,
1998ApJ...501L.189A, 2000ApJ...528..462H, 2006ApJ...640..901H}). Despite
the significant theoretical progress achieved in this field, the details
of this transport mechanism have yet to be fully understood, and it is
not clear if MRI alone is able to account for the mass accretion and
for the removal of excess angular momentum (\citealt{2007MNRAS.376.1740K,
2007A&A...476.1113F, 2008A&A...487....1B}). Given the complexity of this
mechanism and our poor knowledge of its details, the efficiency of angular
momentum transport within the disk is often modeled in the literature
by including a viscosity in the disk modulated via an analogue of the
Shakura-Sunyaev $\alpha$-parameter (\citealt{1973A&A....24..337S}) which
can be expressed in terms of the fluctuating velocity and magnetic
field (\citealt{2002ApJ...578..420R, Romanova2003ApJ, 2008MNRAS.386..673K}).

On the other hand, observations in the X-ray band have shown that
pre-main-sequence stars are strong sources with X-ray luminosity
3-4 orders of magnitude greater than that of the present-day Sun
(\citealt{2005ApJS..160..353G, 2007A&A...468..379A}). The source of
this X-ray radiation is a plasma at $1-100$ MK in the stellar outer
atmospheres (coronae), heated by magnetic activity analogous to the
solar one but much stronger (\citealt{1999ARAA..37..363F}). X-ray
flares are violent manifestations of this magnetic activity and are
triggered by an impulsive energy input from coronal magnetic field. X-ray
observations in the last decades have shown that flares in CTTSs have
amplitudes larger than solar analogues and occur much more frequently
(\citealt{2005ApJS..160..469F, 2005ApJS..160..353G, 2007A&A...468..379A,
2010ApJ...717...93A}). Examples of these flares are those collected by
the Chandra satellite in the Orion star-formation region (COUP enterprise;
\citealt{2005ApJS..160..469F}). The analysis of these flares revealed that
they have peak temperatures often in excess of 100 MK, are long-lasting,
and are confined in very long magnetic structures -- extending for
several stellar radii -- which may connect the star's photosphere with
the accretion disk (structures that could be of the same kind as those
which channel the plasma in the magnetospheric accretion).

At the present time, it is unclear where these flares occur. The
differential rotation of the disk together with the interaction of the
disk with the magnetosphere may cause magnetic reconnection close to
the disk's surface, triggering large-scale flares there. In this case,
the flares may perturb the stability of the circumstellar disk causing, in
particular, a strong local overpressure. The pressure gradient force might
be able to push disk's matter out of the equatorial plane into funnel
streams, thus providing a mechanism to drive mass accretion that differs
from that, commonly invoked in the literature, based on the disk viscosity
which determines the accumulation of disk matter and the increase of gas
pressure close to the truncation radius (\citealt{2002ApJ...578..420R}).
Bright flares close to circumstellar disks may therefore have important
implications for a number of issues such as the transfer of angular
momentum and mass between the star and the disk, the powering of outflows,
and the ionization of circumstellar disks thus influencing also the
efficiency of MRI (see also \citealt{2010ApJ...717...93A}).

In this paper, we investigate the effects of a flare on the stability of
the circumstellar disk with a three-dimensional (3D) magnetohydrodynamic
(MHD) simulation. We model the evolution of the star-disk plasma heated
by a strong energy release (with intensity comparable to that of flares
typically observed in young stars), close to a thick disk surrounding a
rotating magnetized CTTS. The model takes into account all key physical
processes, including the gravitational force, the viscosity of the
disk, the magnetic-field-oriented thermal conduction, the radiative
losses from optically thin plasma, and the coronal heating. The wealth
of nonlinear physical processes governing the star-disk system and the
evolution of the flare made this a challenging task. In Sect. \ref{sec2}
we describe the MHD model and the numerical setup; in Sect. \ref{sec3}
we describe the results; in Sect. \ref{sec4} we discuss the implications
of our results and draw our conclusions.

\section{Problem description and numerical setup}
\label{sec2}

Our model describes a large flare in a rotating magnetized star
surrounded by a thick quasi-Keplerian disk. The flare occurs close
to the inner portion of the disk, within the corotation radius
(i.e. where a Keplerian orbit around the star has the same angular
velocity as the star's surface), where accretion streams are expected
to originate (\citealt{2002ApJ...578..420R, 2008A&A...478..155B}). The
magnetic field of the star is aligned dipole-like, with intensity
$B \approx 1$~kG at the stellar surface according to observations
(\citealt{1999ApJ...510L..41J}). The fluid is assumed to be fully ionized
with a ratio of specific heats $\gamma = 5/3$.

\subsection{MHD equations}

The system is described by the time-dependent MHD equations in a 3D
spherical coordinate system $(R,\theta,\phi)$, extended with gravitational
force, viscosity of the disk, thermal conduction (including the effects
of heat flux saturation), coronal heating (via a phenomenological term),
and radiative losses from optically thin plasma. To our knowledge,
this is the first numerical time-dependent global simulation of the
star-disk system that takes into account simultaneously all key physical
ingredients necessary to describe accurately the effects of a flare on
the structure of the circumstellar disk. The time-dependent MHD equations
written in non-dimensional conservative form are:

\begin{equation}
\frac{\partial \rho}{\partial t} + \nabla \cdot (\rho \vec{u}) = 0~,
\end{equation}

\begin{equation}
\frac{\partial \rho \vec{u}}{\partial t} + \nabla \cdot (\rho
\vec{u}\vec{u}-\vec{B}\vec{B}+\vec{I}P_t-\vec{\tau}) = \rho \vec{g}~,
\end{equation}

\begin{eqnarray}
\lefteqn{\frac{\partial \rho E}{\partial t} +\nabla\cdot [\vec{u}(\rho
E+P_t) -\vec{B}(\vec{u}\cdot \vec{B})-\vec{u}\cdot \vec{\tau}] =} \nonumber \\
 & \displaystyle ~~~~~~~~~~ \rho \vec{u}\cdot
\vec{g} -\nabla\cdot \vec{F}_{\rm c} -n_{\rm e} n_{\rm H}
\Lambda(T)+Q(R,\theta,\phi,t)~,
\end{eqnarray}

\begin{equation}
\frac{\partial \vec{B}}{\partial t} +\nabla
\cdot(\vec{u}\vec{B}-\vec{B}\vec{u}) = 0~,
\end{equation}

\noindent
where

\[
P_t = P + \frac{\vec{B}\cdot\vec{B}}{2}~,~~~~~~~~~~~~~
E = \epsilon +\frac{\vec{u}\cdot\vec{u}}{2}+
\frac{\vec{B}\cdot\vec{B}}{2\rho}~,
\]

\noindent
are the total pressure, and the total gas energy per unit mass (internal
energy, $\epsilon$, kinetic energy, and magnetic energy) respectively, $t$
is the time, $\rho = \mu m_H n_{\rm H}$ is the mass density, $\mu = 1.28$
is the mean atomic mass (assuming metal abundances of 0.5 of the solar
values; \citealt{Anders1989GeCoA}), $m_H$ is the mass of the hydrogen atom,
$n_{\rm H}$ is the hydrogen number density, $\vec{u}$ is the gas velocity,
$\vec{\tau}$ is the viscous stress tensor, $\vec{g}=\nabla\Phi\rs{g}$
is the gravity acceleration vector, $\Phi\rs{g}=- GM_*/R$ is the
gravitational potential of a central star of mass $M_*$, $G$ is the
gravitational constant, $R$ is the distance from the gravitating center,
$T$ is the temperature, $\vec{B}$ is the magnetic field, $\vec{F}_{\rm c}$
is the thermal conductive flux, $\Lambda(T)$ represents the optically thin
radiative losses per unit emission measure derived with the PINTofALE
spectral code (\citealt{Kashyap2000BASI}) and with the APED V1.3 atomic line
database (\citealt{2001ApJ...556L..91S}), assuming the same metal
abundances as before (as deduced from X-ray observations of
CTTSs; \citealt{Telleschi2007A&Ab}), and $Q(R,\theta,\phi,t)$ is a function
of space and time describing the phenomenological heating rate (see
Sect.~\ref{sec:flare}). We use the ideal gas law, $P=(\gamma-1) \rho
\epsilon$.

The viscosity is assumed to be negligible in the extended stellar corona
and effective only in the circumstellar disk. In order to make
the transition between the disk and the corona, we track the original
disk material by using a tracer that is passively advected in the same
manner as the density. We define $C_{\rm disk}$ the mass fraction of
the disk inside the computational cell. The disk material is initialized
with $C_{\rm disk} = 1$, while $C_{\rm disk} = 0$ in the extended corona.
Then the viscosity works only in zones consisting of the original disk
material by more than 99\% or, in other words, where $C_{\rm disk} >
0.99$. The viscous stress tensor is defined as

\begin{equation}
\vec{\tau} = \eta\rs{v}\left[(\nabla \vec{u}) + (\nabla\vec{u})^{\rm
T}-\frac{2}{3}(\nabla\cdot \vec{u})\vec{I}\right]~,
\end{equation}

\noindent
where $\eta\rs{v}=\nu\rs{v}\rho$ is the dynamic viscosity, and
$\nu\rs{v}$ is the kinematic viscosity. Several studies suggest that
turbulent diffusion of magnetic field in the disk is determined by the
same processes that determine turbulent viscosity, leading to angular
momentum transport in the disk (\citealt{2001NewAR..45..663B}). Thus,
we assume that turbulent magnetic diffusivity is of the same order
of magnitude as turbulent viscosity like in the Shakura-Sunyaev model
(\citealt{1973A&A....24..337S}). The kinematic viscosity is expressed
as $\nu\rs{v} = \alpha c\rs{s}^2/\Omega\rs{K}$, where $c\rs{s}$ is
the isothermal sound speed, $\Omega\rs{K}$ is the Keplerian angular
velocity at a given location, and $\alpha< 1$ is a dimensionless parameter
regulating the efficiency of angular momentum transport within the disk,
which can be expressed in terms of the fluctuating velocity and magnetic
field. The parameter $\alpha$ varies in the range $0.01-0.6$ in the
Shakura-Sunyaev accretion model (\citealt{2003ARA&A..41..555B}). In our
simulation, we assume $\alpha = 0.02$.

The thermal conduction is highly anisotropic due to the presence of the
stellar magnetic field, the conductivity being highly reduced in the
direction transverse to the field (\citealt{spi62}). The thermal flux therefore
is locally split into two components, along and across the magnetic field
lines, $\vec{F}_{\rm c} = F_{\parallel}~\vec{i}+F_{\perp}~\vec{j}$.
The thermal conduction formulation also accounts for heat
flux saturation. In fact, during early phases of flares, rapid
transients, fast dynamics and steep thermal gradients are known
to develop (\citealt{2008ApJ...684..715R}). Under these circumstances the
conditions required for classical ``Spitzer'' heat conduction may break
down to the extent that the plasma thermal conduction becomes flux
limited (\citealt{1979ApJ...228..592B}). The two components of thermal flux
are therefore written as (\citealt{2008ApJ...678..274O, 2010A&A...510A..71O})

\begin{equation}
\begin{array}{l}\displaystyle
F_{\parallel} = \left(\frac{1}{[q_{\rm spi}]_{\parallel}}+
                \frac{1}{[q_{\rm sat}]_{\parallel}}\right)^{-1}~,
\\ \\ \displaystyle
F_{\perp} = \left(\frac{1}{[q_{\rm spi}]_{\perp}}+
               \frac{1}{[q_{\rm sat}]_{\perp}}\right)^{-1}~,
\end{array}
\label{cond}
\end{equation}

\noindent
to allow for a smooth transition between the classical and saturated
conduction regime, where $[q_{\rm spi}]_{\parallel}$ and
$[q_{\rm spi}]_{\perp}$ represent the classical conductive flux along
and across the magnetic field lines (\citealt{spi62})

\begin{equation}
\begin{array}{l}\displaystyle
[q_{\rm spi}]_{\parallel} = -\kappa_{\parallel} [\nabla T]_{\parallel}
\approx - 9.2\times 10^{-7} T^{5/2}~ [\nabla T]_{\parallel}
\\ \\ \displaystyle
[q_{\rm spi}]_{\perp} = -\kappa_{\perp} [\nabla T]_{\perp}
\approx - 3.3\times 10^{-16} \frac{n^2_{\rm H}}{T^{1/2}B^2}~ [\nabla
T]_{\perp}~,
\end{array}
\label{spit_eq}
\end{equation}

\noindent
where $[\nabla T]_{\parallel}$ and $[\nabla T]_{\perp}$ are the thermal
gradients along and across the magnetic field, and $\kappa_{\parallel}$
and $\kappa_{\perp}$ (in units of erg s$^{-1}$ K$^{-1}$ cm$^{-1}$)
are the thermal conduction coefficients along and across the magnetic
field, respectively. The saturated flux along and across the magnetic
field lines, $[q_{\rm sat}]_{\parallel}$ and $[q_{\rm sat}]_{\perp}$,
are (\citealt{cm77})

\begin{equation}
\begin{array}{l}\displaystyle
[q_{\rm sat}]_{\parallel} = -\mbox{sign}\left([\nabla
T]_{\parallel}\right)~
                5\varphi \rho c_{\rm s}^3,
\\ \\ \displaystyle
[q_{\rm sat}]_{\perp} = -\mbox{sign}\left([\nabla T]_{\perp}\right)~
                5\varphi \rho c_{\rm s}^3,
\end{array}
\label{therm}
\end{equation}

\noindent
where $\varphi$ is a number of the order of unity; we set
$\varphi = 1$ according to the values suggested for stellar
coronae (\citealt{1984ApJ...277..605G, 1989ApJ...336..979B,
2002A&A...392..735F}).

The calculations were performed using PLUTO
(\citealt{2007ApJS..170..228M}), a modular, Godunov-type code for
astrophysical plasmas. The code provides a multiphysics, multialgorithm
modular environment particularly oriented toward the treatment of
astrophysical flows in the presence of discontinuities as in the case
treated here. The code was designed to make efficient use of massive
parallel computers using the message-passing interface (MPI) library
for interprocessor communications. The MHD equations are solved using
the MHD module available in PLUTO, configured to compute intercell
fluxes with the Harten-Lax-Van Leer approximate Riemann solver, while
second order in time is achieved using a Runge-Kutta scheme. A Van
Leer limiter for the primitive variables is used during the heating
release (at the very beginning of the simulation for $t<300$~s) and a
monotonized central difference limiter at other times.  The evolution
of the magnetic field is carried out adopting the constrained transport
approach (\citealt{1999JCoPh.149..270B}) that maintains the solenoidal
condition ($\nabla\cdot\vec{B}=0$) at machine accuracy. We adopted the
``magnetic field-splitting'' technique (\citealt{1994JCoPh.111..381T,
1999JCoPh.154..284P, 2009A&A...508.1117Z}) by splitting the total
magnetic field into a contribution coming from the background stellar
magnetic field and a deviation from this initial field. Then only the
latter component is computed numerically. This approach is particularly
useful when dealing with low-$\beta$ plasma as it is the case in
proximity of the stellar surface (\citealt{2009A&A...508.1117Z}). PLUTO
includes optically thin radiative losses in a fractional step formalism
(\citealt{2007ApJS..170..228M}), which preserves the $2^{nd}$ time
accuracy, as the advection and source steps are at least of the $2^{nd}$
order accurate; the radiative losses $\Lambda$ values are computed at the
temperature of interest using a table lookup/interpolation method. The
thermal conduction is treated separately from advection terms through
operator splitting. In particular we adopted the super-time-stepping
technique (\citealt{sts}) which has been proved to be very effective
to speed up explicit time-stepping schemes for parabolic problems. This
approach is crucial when high values of plasma temperature are reached
(as during flares), explicit scheme being subject to a rather restrictive
stability condition (i.e. $\Delta t < (\Delta x)^2/(2\eta)$, where $\eta$
is the maximum diffusion coefficient), as the thermal conduction timescale
$\tau\rs{cond}$ is shorter than the dynamical one $\tau\rs{dyn}$ (e.g.
\citealt{2000A&A...362L..41H, 2005CoPhC.168....1H, 2005A&A...444..505O,
2008ApJ...678..274O}); in particular, during the early phases of the
flare evolution, we find $\tau\rs{cond}/\tau\rs{dyn} \approx 10^{-2}$. The
viscosity is solved with an explicit scheme, using a second-order finite
difference approximation for the dissipative fluxes.

\subsection{Initial and boundary conditions}
\label{sec:inicond}

A star of mass $M_* = 0.8 M_{\odot}$ and radius $R_* = 2 R_{\odot}$
is located at the origin of the 3D spherical coordinate system
$(R,\theta,\phi)$, with the rotation axis coincident with the normal
to the disk midplane. The rotation period of the star is assumed to be
9.2 days. The initial unperturbed stellar atmosphere is approximately in
equilibrium and consists of three components: the stellar magnetosphere,
the extended stellar corona, and the quasi-Keplerian disk.

The pre-flare magnetosphere is assumed to be force-free, with dipole
topology and magnetic moment $\mu\rs{B}$ aligned with the rotation axis
of the star. Thus the two field components in spherical coordinates are

\begin{equation}
B\rs{R} = \frac{2\mu\rs{B}\cos\theta}{R^3}~,~~~~~~~
B\rs{\theta} = \frac{\mu\rs{B}\sin\theta}{R^3}~.
\end{equation}

\noindent
The magnetic moment is chosen in order to have a magnetic field
strength of the order of 1~kG at the stellar surface according
to observations (\citealt{1999ApJ...510L..41J}).

The initial corona and disk are set in order to satisfy mechanical
equilibrium involving centrifugal, gravitational, and pressure
gradient forces. In particular, we adopted the initial conditions
introduced by \cite{2002ApJ...578..420R} (where the reader is referred
to for more details) and describing the star-disk system in quiescent
configuration. These conditions assume that initially the plasma
is barotropic and that the disk and the corona are both isothermal
with temperatures $T\rs{d}$ and $T\rs{c}$, respectively. With these
assumptions, the thermal pressure at any point of the spatial domain is
given by

\begin{equation}
P = \left\{
\begin{array}{ll}
\displaystyle P_0\exp\left[{\cal F}\frac{\mu m\rs{H}}{2T\rs{c}}\right] & ~~~P\leq P_0~,\\
\displaystyle P_0\exp\left[{\cal F}\frac{\mu m\rs{H}}{2T\rs{d}}\right] & ~~~P\geq P_0~,\\
\end{array}\right.
\label{ini_pres}
\end{equation}

\noindent
where $P_0 = 2\rho\rs{c}k\rs{B}T\rs{c}/(\mu m\rs{H})$ is the initial
pressure at the boundary between the disk and the corona, $\rho\rs{c}$
is the mass density of the corona close to the disk truncation radius
$R\rs{d}$, $k\rs{B}$ is the Boltzmann constant, ${\cal F}$ is the function

\begin{equation}
{\cal F} = (k-1)\frac{GM_*}{R\rs{d}}-(\Phi\rs{g}+\Phi\rs{c})~,
\end{equation}

\noindent
$k$ is a constant of the order of 1 that takes into account
that the disk is slightly non-Keplerian (we set $k = 1.01$ as in
\citealt{2002ApJ...578..420R}), $\Phi\rs{c}$ is the centrifugal potential
written as

\begin{equation}
\Phi\rs{c} = \left\{
\begin{array}{ll}
\displaystyle k\frac{GM_*}{R\rs{d}}\left[1+\frac{R\rs{d}^2-r^2}{2R\rs{d}^2}\right] & ~~~r
\leq R\rs{d}~,\\
\displaystyle k\frac{GM_*}{r} & ~~~r \geq R\rs{d}~,
\end{array}\right.
\end{equation}

\noindent
and $r = R\sin\theta$ is the cylindrical radius. The mass density depends
on the pressure given in Eq.~\ref{ini_pres} at any point of the spatial domain

\begin{equation}
\rho(P) = \left\{
\begin{array}{ll}
\displaystyle \frac{\mu m\rs{H} P}{2 T\rs{c}} & ~~~P < P_0~,\\
\displaystyle \frac{\mu m\rs{H} P}{2 T\rs{d}} & ~~~P > P_0~.
\end{array}\right.
\end{equation}

\noindent
The angular velocity of the plasma is given by

\begin{equation}
\omega = \left\{
\begin{array}{ll}
\displaystyle \left(k \frac{GM_*}{R\rs{d}^3}\right)^{1/2} & ~~~r\leq R\rs{d}~, \\
\displaystyle \left(k \frac{GM_*}{r^3}\right)^{1/2} & ~~~r\geq R\rs{d}~.
\end{array}\right.
\label{ini_omega}
\end{equation}

In our simulation, we assume the isothermal disk to be cold ($T\rs{d}
= 8\times 10^3$ K), dense ($n\rs{d}$ ranges between $5\times 10^{11}$
and $4\times 10^{12}$ cm$^{-3}$), and to rotate with angular velocity
close to the Keplerian value $\Omega\rs{K}$. The rotation axis of the
disk (coincident with the rotation axis of the star) is aligned with
the magnetic moment $\mu\rs{B}$. The disk is initially truncated by the
stellar magnetosphere at the radius $R\rs{d}$ where the total gas pressure
of the disk equals the magnetic pressure, $p+\rho u^2 = B^2/8\pi$; for
the parameters adopted here, $R\rs{d} = 2.86\, R_*$. In our simulation,
the corotation radius $R\rs{co} = (GM_*/\Omega_*^2)^{1/3}=9.2\, R_*$,
where $\Omega_*$ is the stellar angular velocity. The corona is initially
isothermal with $T\rs{c} = 4$ MK and at low density\footnote{Note however
that the density of the outer corona is not important for the flare
evolution.} with $n\rs{c}$ ranging between $\approx 10^8$ and $10^9$
cm$^{-3}$. As shown in Eq.~\ref{ini_omega}, we allow the corona to be
initially rotating with angular velocity equal to the Keplerian rotation
rate of the disk in order to have approximately equilibrium conditions
and reduce the effects of transients caused by the initial differential
rotation between the disk and the corona (\citealt{2002ApJ...578..420R,
2009A&A...508.1117Z}).

\begin{figure}
  \centering
  \includegraphics[width=7truecm]{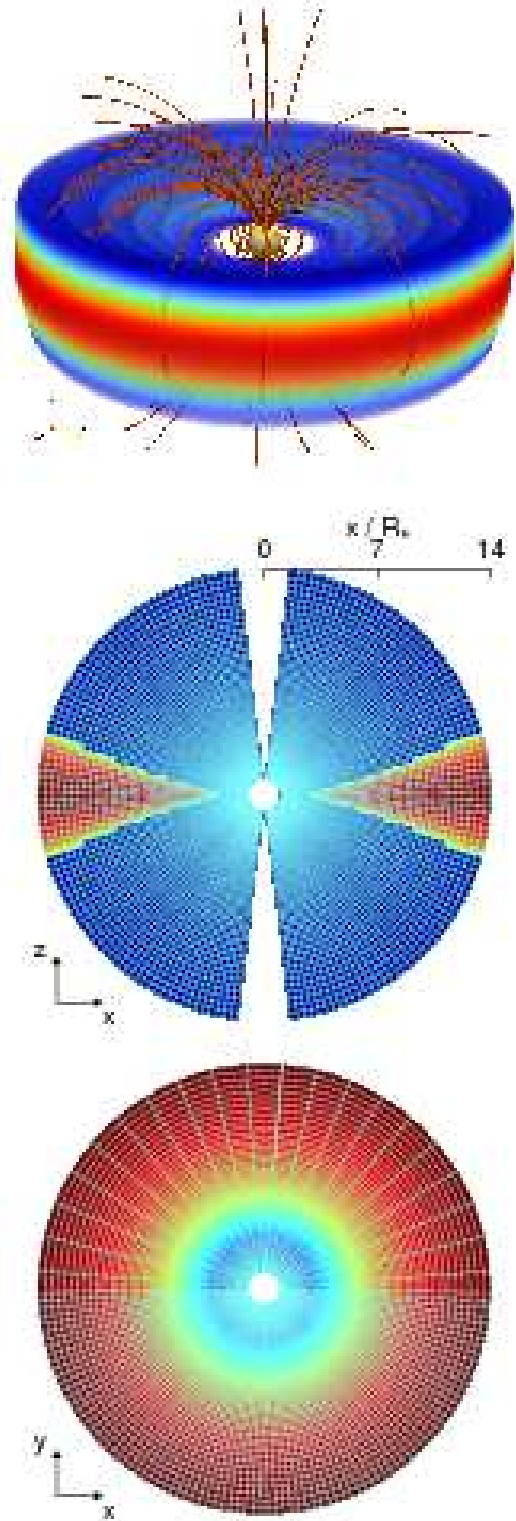}
  \caption{Initial conditions and computational domain. {\em Upper panel}:
           volume rendering of the mass density, in log scale, at the
           beginning of the simulation ($t=0$). The selected magnetic
           field lines marked in red describe the initial dipolar magnetic
           field. The yellow sphere at the center of the spatial domain
           represents the central protostar. {\em Middle panel}: slice
           in the $(x,z)$ plane of the mass density distribution with
           overplotted the computational grid. {\em Bottom panel}: as
           in the middle panel for the slice in the equatorial plane
           $(x,y)$.}
  \label{fig1_sup}
\end{figure}

Figure~\ref{fig1_sup} shows the initial condition together with the
numerical grid adopted in our simulation. The computational domain
extends between $R\rs{min} = R_*$ (i.e. the inner boundary coincides with
the stellar surface) and $R\rs{max} = 14\, R_*$ in the radial direction,
and encompasses an angular sector going from $\theta\rs{min} = 5^0$ to
$\theta\rs{max} = 175^0$ in the angular coordinate $\theta$, and from
$\phi\rs{min} = 0^0$ to $\phi\rs{max} = 360^0$ in the angular coordinate
$\phi$. The inner and outer boundaries in $\theta$ do not coincide
with the rotation axis of the star-disk system to avoid extremely small
$\delta \phi$ values, vastly increasing the computational cost. On the
other hand, all the evolution relevant for this study never involve
portions of the domain close to the star-disk rotation axis.

The radial coordinate $R$ has been discretized on a logarithmic grid
with the mesh size increasing with $R$ (see Fig.~\ref{fig1_sup}), giving
a higher spatial resolution closer to the star as it is appropriate
for simulations of accretion flows to a star with a dipole field
(\citealt{2002ApJ...578..420R}). The radial grid is made of $N\rs{R}
= 80$ points with a maximum resolution of $\Delta R = 4.8\times 10^9$
cm close to the star and a minimum resolution of $\Delta R = 6.4\times
10^{10}$ cm close to the outer boundary. The angular coordinate $\theta$
has been discretized uniformly with $N\rs{\theta} = 90$ points, giving
a resolution of $\Delta \theta = 2^o$. The angular coordinate $\phi$ is
nonuniform with the highest resolution in an angular sector of $180^0$
placed where the flaring loop and the stream evolve (see bottom panel
in Fig.~\ref{fig1_sup}). The $\phi$-grid is made of $N\rs{\phi} = 110$
points with a maximum resolution of $\Delta \phi = 2^o$ and a minimum
resolution of $9^0$. The numerical grid is not static but tracks the
hot loop and the stream as the calculation progresses, in such a way
that the loop and the stream evolve in the portion of the domain with
the highest spatial resolution (namely that with $\Delta \phi = 2^o$).

The boundary conditions at the stellar surface $R\rs{min}$ amount to
assuming that the infalling material passes through the surface of the
star as done by \cite{2002ApJ...578..420R} (outflows boundary condition;
see also \citealt{Romanova2003ApJ, 2008MNRAS.386..673K}), thus ignoring
the dynamics of the plasma after it impacts on the star. Zero-gradient
boundary conditions are assumed at the outer boundary of the coordinate
$R$ ($R\rs{max}$) and at the boundaries of the angular coordinate $\theta$
($\theta\rs{min}$ and $\theta\rs{max}$). Finally, periodic boundary
conditions are assumed for angular coordinate $\phi$ ($\phi\rs{min}$
and $\phi\rs{max}$).

\subsection{Coronal heating and flare}
\label{sec:flare}

The phenomenological heating is prescribed as a component, describing
the stationary coronal heating, plus a transient component, triggering
a flare. The former component works at temperature $T \leq 4$ MK and
is chosen to balance exactly the local radiative losses, leading to
a quasi-stationary extended corona during the whole simulation. The
transient component describes the injection of a heating pulse at the
surface of the accretion disk. The flare could be driven by a stressed
field configuration resulting from the twisting of magnetic field lines
induced by the differential rotation of the inner rim of the disk and the
stellar photosphere (\citealt{1997Sci...277.1475S}). The heat pulse has a 3D
Gaussian spatial distribution located at the disk border at a distance
of 5 $R_*$ (namely well below the corotation radius $R\rs{co}= 9.2\,
R_*$) with width $\sigma = 2\times 10^{10}$ cm. Its intensity per unit
volume is $H_0 = 32$ ergs cm$^{-3}$ s$^{-1}$. The pulse starts at the
beginning of the simulation ($t = 0$) and is switched off completely
after 300 s. The flare parameters are analogous to those adopted
in a one-dimensional (1D) hydrodynamic model that reproduces the
evolution of a large flare observed in an Orion young star from COUP
(\citealt{2005ApJS..160..469F}). The total energy released during our
simulated flare is $E\rs{fl} = 10^{36}$ ergs, namely the same order of
magnitude of the energy involved in the brightest X-ray flares observed
in COUP (\citealt{2005ApJS..160..469F, 2005ApJS..160..423W}).

\section{Results}
\label{sec3}

\begin{figure*}
  \centering
  \includegraphics[width=17.5truecm]{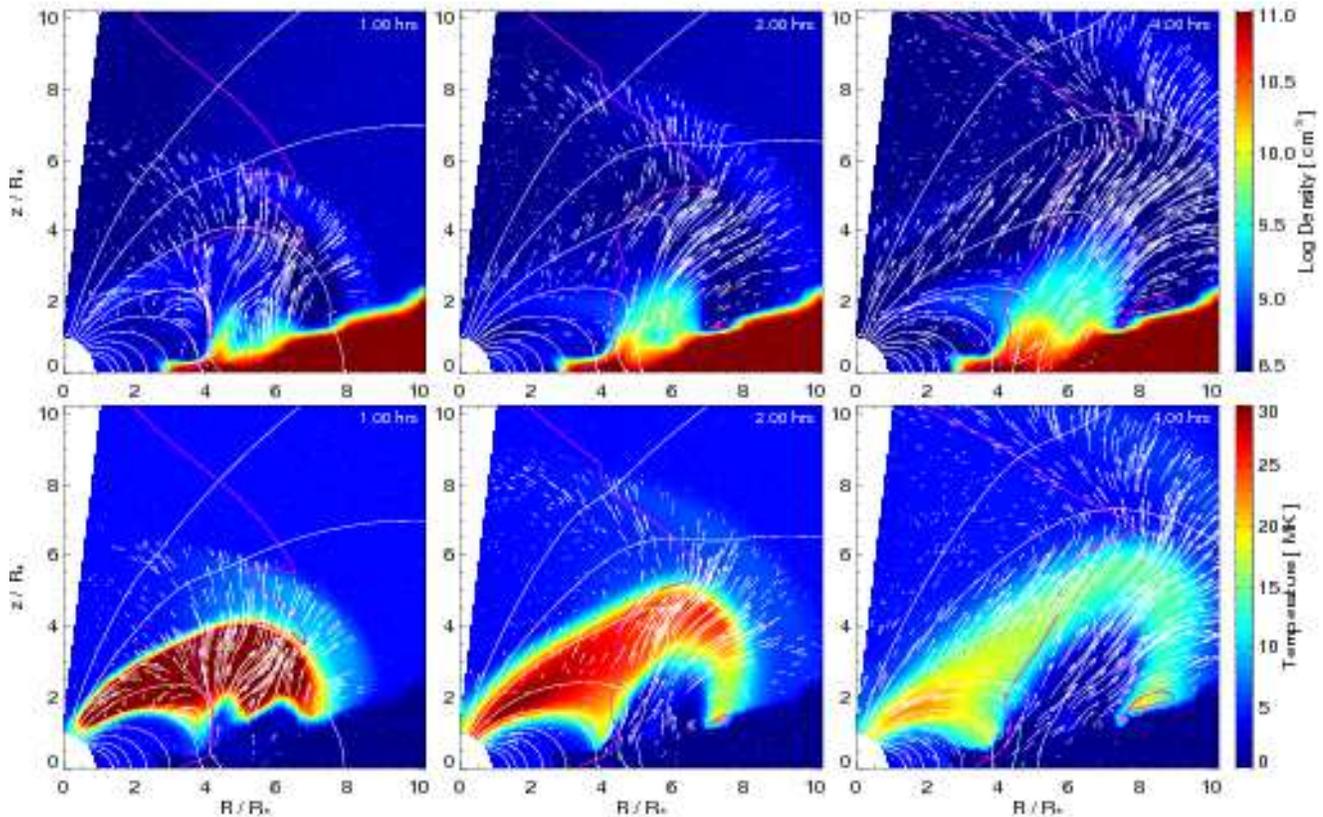}
  \caption{Close view of the flaring loop during its early
           evolution. The figure shows slices in the $(R,z)$ plane
           passing through the middle of the flaring loop, reporting
           the distributions of density (upper panels) and temperature
           (lower panels). The slices encompass an angular sector
           going from the rotation axis to the disk midplane. The white
           lines represent sampled poloidal magnetic field lines. The
           arrows represent the poloidal flow velocities. The magenta line
           delimits the region with plasma $\beta < 1$ (on the left of
           each panel).}
     \label{flare_evol}
\end{figure*}

\subsection{Evolution of the flare and X-ray emission}
\label{evol_flare}

\begin{figure*}
  \centering
  \includegraphics[width=17.5truecm]{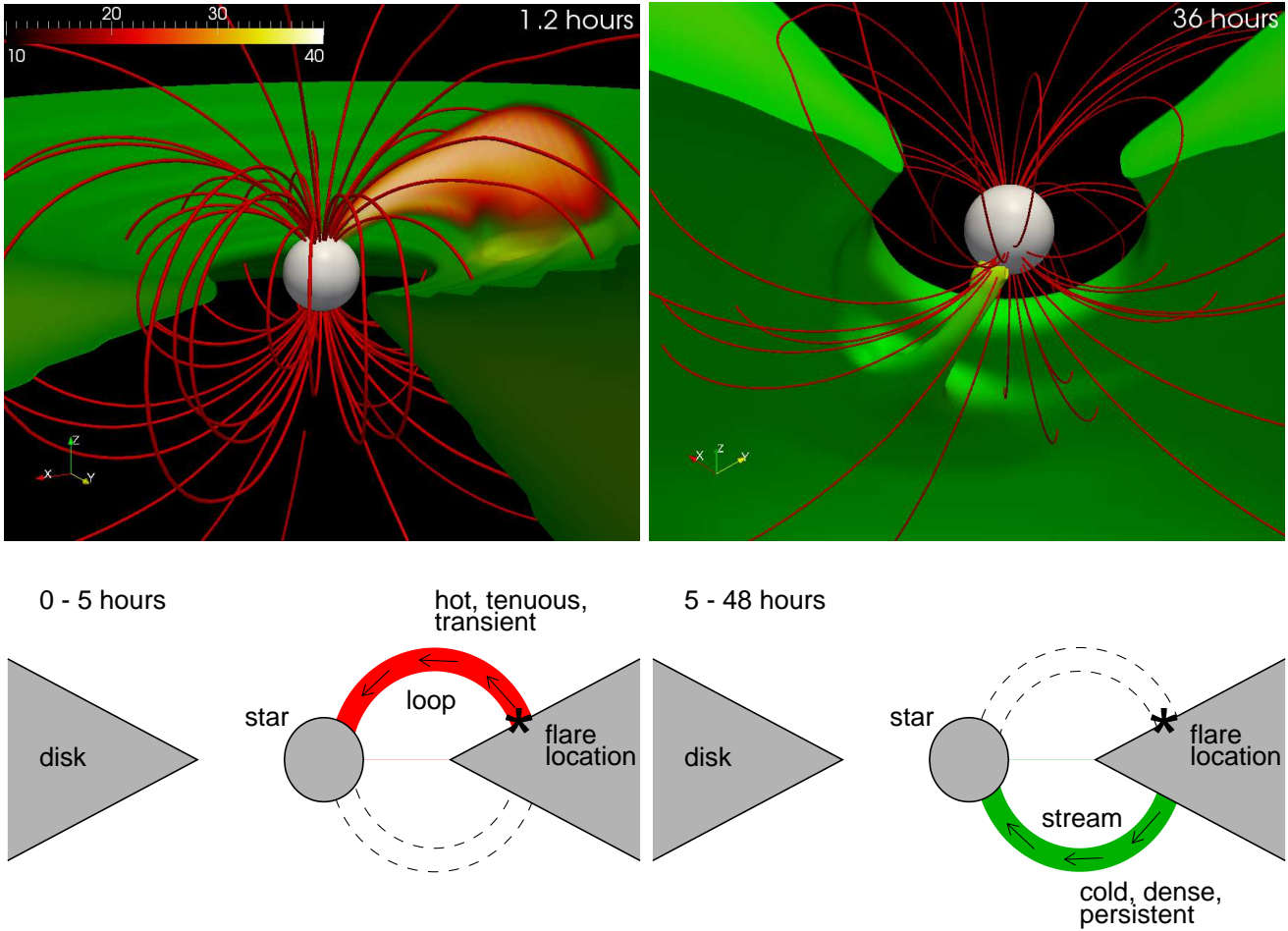}
  \caption{Evolution of the bright flare and accretion stream. The upper
           panels show cutaway views of the star-disk system showing the
           mass density of the disk (green) at $t=1.2$ hours (on the left)
           and, from the opposite side, at $t=36$ hours (on the right)
           since the injection of the heat pulse. The upper left panel
           also overplots the three-dimensional volume rendering of
           the plasma temperature (in MK; see color table in the upper
           left corner of the panel), showing the flaring loop (in red)
           linking the inner part of the disk with the star. The upper
           right panel shows the accretion stream triggered by the flare
           in the side of the disk opposite to the flaring loop. Selected
           magnetic field lines are overplotted in red. Lower panels
           show schematic views of the system during the evolution of
           the flaring loop (on the left) and in the subsequent period,
           during the evolution of the stream (on the right).}
      \label{fig1}
\end{figure*}

We followed the evolution of the star-disk system for $\approx 2$ days,
focusing on its effects on the disk structure. Figure~\ref{flare_evol}
shows the distributions of density and temperature in $(R,z)$
slices passing through the middle of the heating pulse during its
evolution. During this impulsive phase, the local magnetic field is
perturbed by the flare and an MHD shock wave develops in the magnetosphere
above the disk and propagates radially away from the central protostar
in regions where $\beta > 1$. At the same time, the sudden heat
deposition determines a local increase of temperature (up to a maximum of
800 MK) and pressure (above 6000 dyn cm$^{-2}$). The dense disk material
is heated and expands in the magnetosphere with a strong evaporation front
at speeds above 4000 km s$^{-1}$. The fast thermal front propagates toward
the star along the magnetic field lines and reaches the stellar surface on
a timescale of $\approx 1$ hour (left panels in Fig.~\ref{flare_evol}). At
this point, a hot magnetic tube (loop) of length $L \approx 10^{12}$~cm
(i.e. comparable to loop lengths inferred from COUP observations;
\citealt{2005ApJS..160..469F}) is formed, linking the inner part of
the disk to the star's photosphere. The loop is illustrated in the left
panels of Fig.~\ref{fig1} showing a cutaway view of the star-disk system
at $t=1.2$ hours (upper panel; the flaring loop is marked in red), and
a schematic view of the system during the evolution of the flaring loop
(lower panel). Due to the efficient thermal conduction and radiation,
the plasma begins to cool immediately after the heat pulse is over, and
the maximum temperature rapidly decreases to $\sim 50$ MK in $\approx 1$
hour. The disk evaporation is fed by thermal conduction from the outer
hot plasma, even after the end of the heat pulse.

Figure~\ref{flare_evol} also shows that the overheating of the disk
surface at the loop footpoint makes a significant amount of material
expand and be ejected in the magnetosphere. A small fraction of this
evaporated disk material streams along the loop accelerated toward the
star by its gravity, fills the whole tube in $\approx 10$ hours, and the
density keeps increasing throughout the loop up to values ranging between
$10^{9}$ and $10^{10}$ cm$^{-3}$ at its apex.  On the other hand, most of
the evaporated disk material is not efficiently confined by the magnetic
field and channeled into the hot loop but is rather ejected away from
the central star in the outer stellar corona, carrying away mass and
angular momentum (see Fig.~\ref{flare_evol}).  Due to the high values of
$\beta$ there, the magnetic field lines are dragged away. The outflowing
plasma is supersonic and its speed is of the order of the local Alfv\'en
speed $u\rs{A} \approx 300$ km s$^{-1}$. We note that our results on
the dynamics of the outflowing plasma are similar to those found with
an MHD model proposed by \cite{1996ApJ...468L..37H} to describe hard
X-ray flares in protostars observed by the ASCA satellite.

From the model results, we synthesized the X-ray emission originating
from the flare, applying a methodology analogous to that described
in the literature in the context of the study of novae and supernova
remnants (\citealt{orlando2, 2009A&A...493.1049O}). In particular,
we first calculate the emission measure in the $j$-th domain cell
as ${\rm em}\rs{j} = n\rs{Hj}^2 V\rs{j}$ (where $n\rs{Hj}^2$ is the
hydrogen number density in the cell, $V\rs{j}$ is the cell volume, and
we assume fully ionized plasma). From the values of emission measure and
temperature in the cell, we synthesize the corresponding X-ray spectrum,
using the PINTofALE spectral code (\citealt{Kashyap2000BASI}) with the
APED V1.3 atomic line database, and assuming the same metal abundances
of the simulation, namely 0.5 of the solar values, as deduced from
X-ray observations of CTTSs (\citealt{Telleschi2007A&Ab}). We integrate
the X-ray spectra from the cells in the whole spatial domain and, then,
derive the X-ray luminosity by integrating the spectrum in the $[0.6-12]$
keV band.

\begin{figure}
  \centering
  \includegraphics[width=8.3truecm]{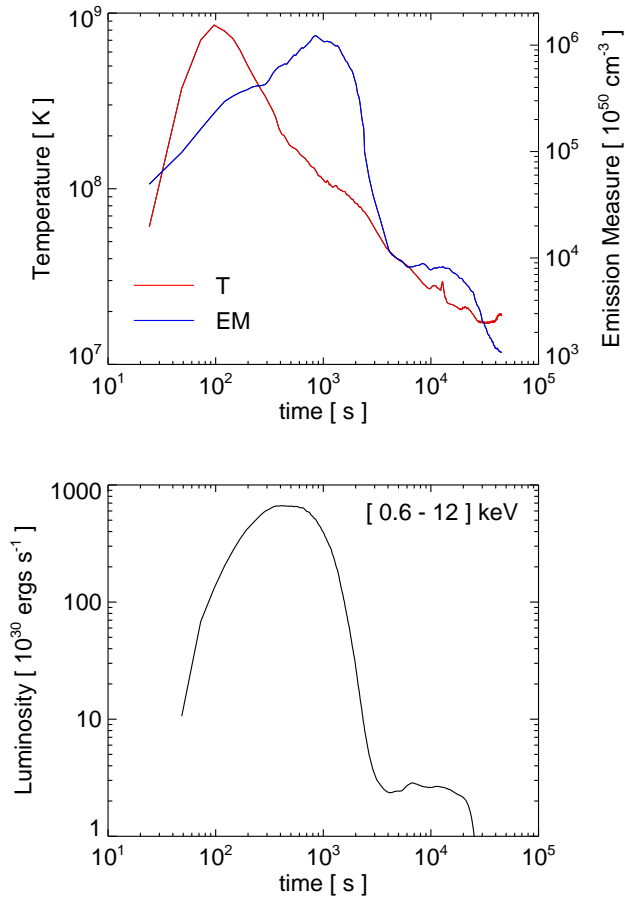}
  \caption{{\em Upper panel}: evolution of maximum temperature (red line)
           and emission measure (blue line) of the flaring loop. {\em
           Lower panel}: X-ray lightcurve of the flare in the $[0.6-12]$
           keV band.}
  \label{fig2_sup}
\end{figure}

Figure~\ref{fig2_sup} shows the evolution of the maximum temperature
and emission measure of the flaring loop, and its X-ray lightcurve. The
peak temperature of the flare is reached very soon at $t\approx 100$~s,
the emission measure peaks later at $t\approx 1000$~s. The
X-ray luminosity evolves as the emission measure, reaching a peak
value of $L\rs{X} \approx 6.5\times 10^{32}$~ergs~s$^{-1}$ at $t\approx
500$~s, namely after the end of the heat pulse. Then the luminosity
(and the emission measure) decreases by more than two orders of
magnitude until $t\approx 3\times 10^3$~s, is steady till $t\approx
2\times 10^4$~s, and decreases again afterward. The peak
X-ray luminosity of the simulated flare is consistent with the values
derived for the brightest X-ray flares observed in COUP (see Table 1 in
\citealt{2005ApJS..160..469F}) which range between $10^{32}$~ergs~s$^{-1}$
(source COUP 752) and $8\times 10^{32}$~ergs~s$^{-1}$ (source COUP
1568). On the other hand, it turns out that 3/4 of the COUP flares have a
peak X-ray luminosity approximately one order of magnitude lower than that
simulated here although the total energy released during the simulated
flare (namely $E\rs{fl} = 10^{36}$ ergs, see Sect.~\ref{sec:flare})
is comparable with the median total energy of the flare inferred from
the COUP observations (\citealt{2005ApJS..160..423W}). This may be
due to the fact that the simulated flare lasts for a time interval
significantly shorter than those typical of stellar flares and is only
partially confined by the magnetic field (see discussion below).

The evolution of our simulated flare has significant differences
with respect to the evolution of flares simulated with 1D models
(\citealt{1988ApJ...328..256R}). In fact, at variance with 1D models
where the flare is assumed to be fully confined by the magnetic field,
in our simulation the flare is only partially confined by the magnetic
field at the loop footpoint anchored at the disk surface. There, a
substantial amount of the evaporated disk material escapes in the outer
stellar magnetosphere and does not fill the post-flare loop. We conclude
therefore that our results could be intermediate between those found
with models of fully confined flares (\citealt{1988ApJ...328..256R})
and those of models of unconfined flares (\citealt{2002A&A...383..952R})
which show that the the flare evolution is much faster than that observed
in the confined case.  On the other hand, for the purposes of this work
(namely the study of the perturbation of the disk by a bright flare), we
have used a simplified and idealized configuration of the initial stellar
magnetic field (namely a magnetic dipole), whilst many observations
indicate that the stellar atmospheres are permeated by magnetic fields
with a high degree of complexity (e.g. \citealt{2010RPPh...73l6901G} and
references therein). In particular, the magnetic field in proximity of
a heat release (due to magnetic reconnection) near the disk is expected
to be more complex than that modeled here. In the presence of complex
magnetic field configurations, the magnetic structure hosting the flaring
plasma is expected to confine more efficiently the hot plasma, producing
a flare evolution more similar to that described by 1D models.

It is worth emphasizing that our 3D simulation is focused on the effects
of the flare on the stability of the disk and does not pretend to
describe accurately the evolution of the flaring loop. Nevertheless, we
show here that, even if the flare evolution is not described accurately,
the length, maximum temperature, and peak X-ray luminosity of the flaring
loop reproduced by our simulation resemble those derived from the analysis
of the brightest X-ray flares observed in young low-mass stars
(\citealt{2005ApJS..160..469F}). We are confident, therefore, that the
flare simulated here is appropriate to investigate the effects of bright
flares observed in young stars on the stability of the disk.

\subsection{Dynamics of the accretion stream}

\begin{figure*}
  \centering
  \includegraphics[width=17.5truecm]{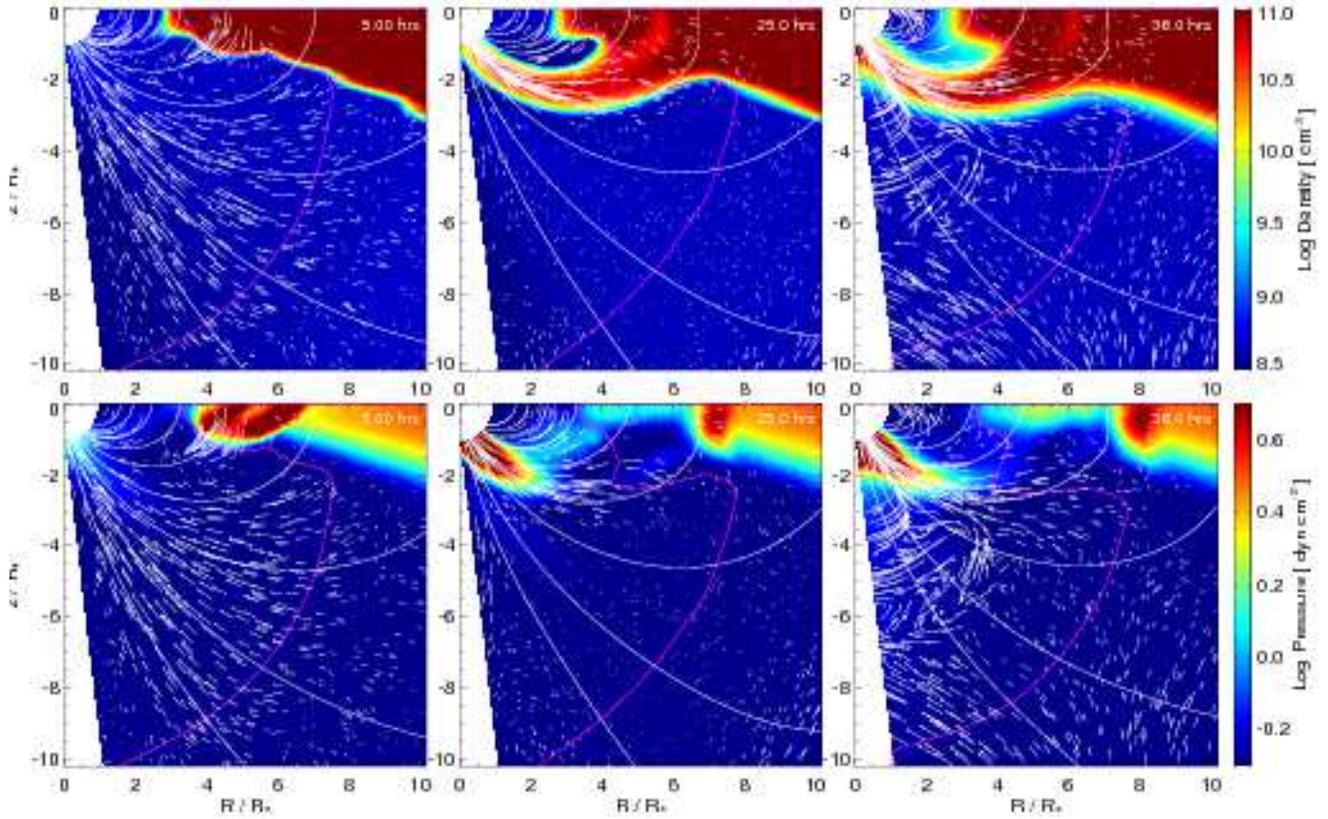}
  \caption{Close view of the accretion stream during its evolution.
           The figure shows slices in the $(R,z)$ plane passing through
           the middle of the stream, reporting the distributions of
           density (upper panels) and pressure (lower panels). The
           slices encompass an angular sector going from the disk
           midplane to the rotation axis. The white lines represent
           sampled poloidal magnetic field lines. The arrows represent
           the poloidal flow velocities. The magenta line delimits the
           region with plasma $\beta < 1$ (on the left of each panel).}
  \label{stream_evol}
\end{figure*}

During the flare evolution, the injected heat pulse produces an
overpressure in the disk at the footpoint of the loop. This overpressure
travels through the disk and reaches the opposite boundary after $\approx
5$ hours, where it pushes the plasma out to form an intense funnel
stream. Figure~\ref{stream_evol} shows the distributions of density and
pressure in $(R,z)$ slices passing through the middle of the stream. The
overpressure wave triggering the stream is evident in the bottom panels.
This new intense stream flows along the dipolar magnetic field lines and
impacts onto the stellar surface $\approx 25$ hours after the injection
of the heat pulse. The right panels in Fig.~\ref{fig1} show a
cutaway view of the star-disk system (upper panel) after the impact of the
stream onto the stellar surface and a schematic view
of the system during the stream evolution (lower panel). As a result,
the stream accretes substantial mass onto the young star from the side
of the disk opposite to the post-flare loop. Our 3D simulation follows
the evolution of the accretion stream for additional 23 hours, for a
total of 48 hours. In this time lapse the stream gradually approaches
a quasi-stationary condition.

\begin{figure*}
  \centering
  \includegraphics[width=16truecm]{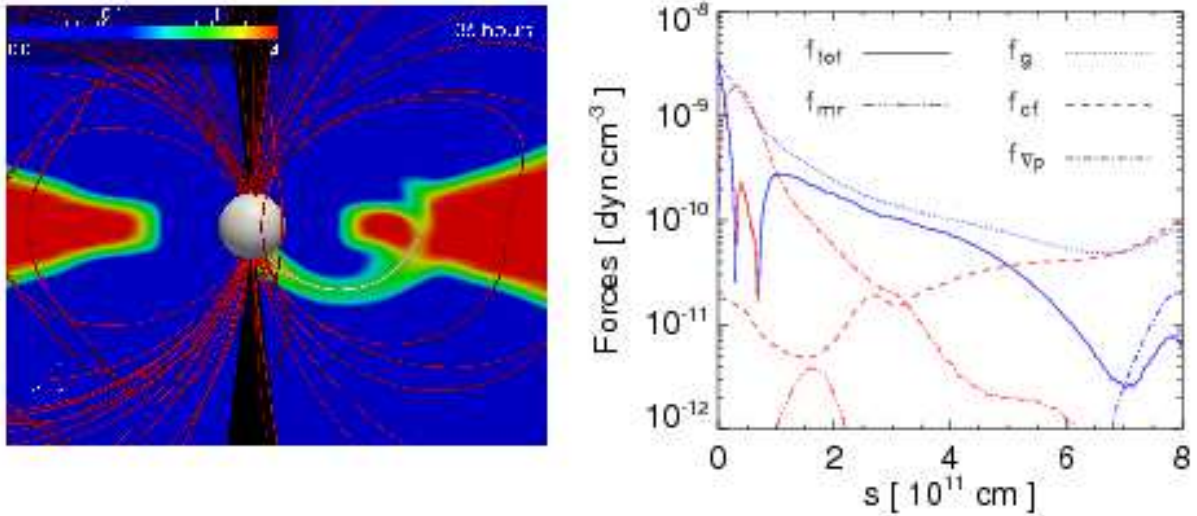}
  \caption{{\em Left panel}: two-dimensional section, in the $(x,z)$
           plane, of the particle number density distribution, in
           log scale, in units of $10^{11}$~cm$^{-3}$ (see color
           table in the upper left corner of the panel) at the time
           $t=36$~hours. The slice passes through the middle of
           the accretion stream. Selected magnetic field lines are
           overplotted in red; the white line represents a
           segment of a fiducial magnetic field line approximately
           in the middle of the accretion stream. {\em Right panel}:
           forces governing the stream dynamics versus the distance from
           the stellar surface along the fiducial field line in the left
           panel: $f\rs{mr}$ is the backward force due to the magnetic
           mirror effect, $f\rs{g}$ the gravitational force, $f\rs{cf}$
           the centrifugal force, $f\rs{\nabla p}$ the pressure gradient
           force, and $f\rs{tot}$ is the resultant force. Forces pushing
           the matter toward (away from) the star are in blue (red).}
  \label{fig2}
\end{figure*}

We analyzed the dynamics of the stream by deriving the forces at work
along a fiducial magnetic field line nested within the stream. Initially
the stream is triggered by a strong pressure gradient due to the
overpressure wave originating from the flare. The pressure gradient
force drives the material out of the disk and channels it into a funnel
flow. Then the gravitational force accelerates the escaped material toward
the central star. These forces evolve with time and Fig.~\ref{fig2}
shows them along the fiducial field line after the stream impacts
the stellar surface. At this stage, the pressure gradient is still
effective in pushing the disk material out of the disk, but it acts
in the opposite direction (against the free-fall of matter) in most of
the accretion stream. The pressure gradient becomes the dominant force
close to the stellar surface, substantially braking (but not stopping) the
accretion flow. As discussed below, at this stage of evolution, the stream
has not yet reached a quasi-stationary condition. Later, when the stream
stabilizes, the gravitational force dominates the stream evolution and the
plasma continuously accelerates toward the star, approaching the free-fall
speed $u\rs{ff}$ at the star's photosphere. Other forces acting against
the free-fall of matter are the centrifugal force and a backward force
due to the magnetic mirror effect when the material approaches the star
(see also \citealt{2002ApJ...578..420R, 2009A&A...508.1117Z}). However,
these forces are much smaller than the others and do not play any
relevant role in the stream dynamics (Fig.~\ref{fig2}; see also
\citealt{2002ApJ...578..420R, 2009A&A...508.1117Z}).

\begin{figure}
  \centering
  \includegraphics[width=7.5truecm]{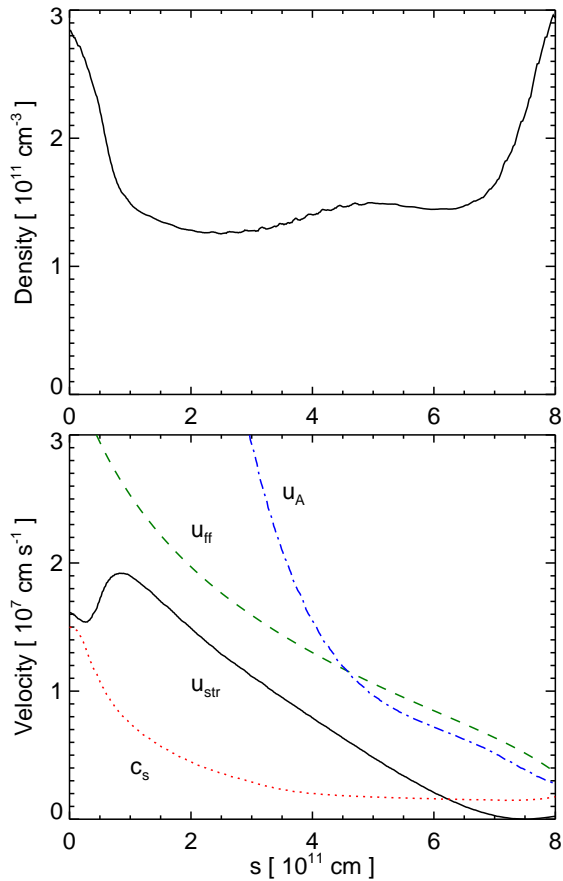}
  \caption{Profiles of particle number density (upper panel) and relevant
           velocities (lower panel) versus the distance $s$ from the
           stellar surface along the fiducial magnetic field line shown
           in Fig.~\ref{fig2} at time $t=36$~hours: $u\rs{str}$ is the
           velocity of matter flowing along the accretion stream (solid
           black line), $c\rs{s}$ the isothermal sound speed (dotted red
           line), $u\rs{ff}$ the free-fall velocity (dashed green line),
           and $u\rs{A}$ the Alfv\'en speed (dot-dashed blue line).}
  \label{fig4_s}
\end{figure}

Figure~\ref{fig4_s} shows the profiles of particle number density
and different velocities along the fiducial magnetic field line
shown in Fig.~\ref{fig2} at time $t=36$~hours. The matter flows with
poloidal velocity $u\rs{str}$. The flow is accelerated by gravity
and becomes supersonic at a distance of $\approx 2\times 10^{11}$
cm from the disk, while the stream density decreases. The flow
gradually approaches the free-fall velocity $u\rs{ff}$, reaching a
maximum velocity of $u\rs{str}\approx 200$~km s$^{-1}$ at $s\approx
10^{11}$~cm ($u\rs{str}\approx 0.8 u\rs{ff}$). Then the flow slightly
brakes while the stream density increases again approaching the stellar
surface. As discussed above, the slowdown of the flow is due to the
pressure gradient force acting against the free-fall of matter for $s$
ranging between $0.5\times 10^{11}$ and $1\times 10^{11}$~cm (see
Fig.~\ref{fig2}). This feature is present until the end of our 3D
simulation at $t\approx 48$~hours.

\begin{figure}
  \centering
  \includegraphics[width=7.5truecm]{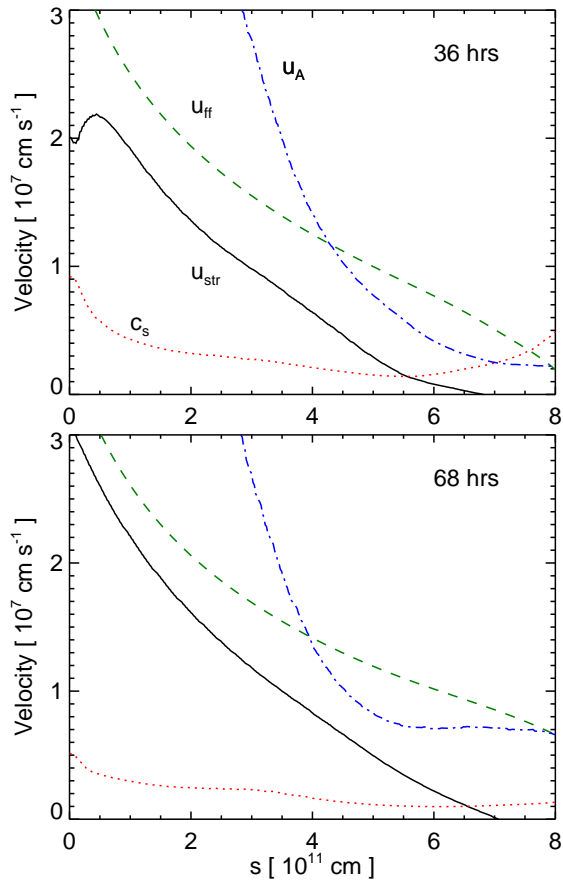}
  \caption{Profiles of velocity along the stream in the 2.5 D
           simulation. As in Fig.~\ref{fig4_s} for the profiles of
           relevant velocities with the distance $s$ from the stellar
           surface along a fiducial magnetic field line nested within the
           stream at time $t=36$~hours (upper panel) and $t=68$~hours
           (lower panel).}
  \label{stream_prof_2D}
\end{figure}

To further investigate the evolution of the accretion stream, we
performed an additional simulation with the same parameters of the 3D
simulation discussed above but carried out in 2.5 dimensions (2.5D),
that is, in spherical coordinates $(R,\theta)$ assuming axisymmetry
around the rotation axis of the star-disk system. Note that, in this
configuration, the heat pulse is not localized in a relatively small
portion of the disk (as in our 3D simulation), but is distributed in
a ring. Nevertheless, we found that the evolution of the flare and the
stream described by the 2.5D simulation is quite similar to that of our
3D simulation. The 2.5D simulation allowed us to extend our analysis
of the stream dynamics, following its evolution until $t= 100$~hours
(i.e. approximately four days). Figure~\ref{stream_prof_2D} shows the
profiles of the relevant velocities derived from the 2.5D simulation at
$t=36$~hours (upper panel) and $t=68$~hours (lower panel). The velocity
profiles at $t=36$~hours resemble those derived from the 3D simulation
(lower panel in Fig.~\ref{fig4_s}). In particular, the flow slightly
brakes approaching the stellar surface due to a pressure gradient force
slowing down the free-fall of matter, as found in the 3D simulation. In
addition, the 2.5D simulation shows that, at this stage, the stream has
not reached yet a quasi-stationary condition. Later, the stream stabilizes
(after $t\approx 60$ hours); the gravitational force becomes dominant
along the stream and the matter is accelerated until it impacts the
stellar surface, reaching there a maximum velocity of $u\rs{str}\approx
300$~km~s$^{-1}$ corresponding to $\approx 0.9\, u\rs{ff}$.

\begin{figure}
  \centering
  \includegraphics[width=7.8truecm]{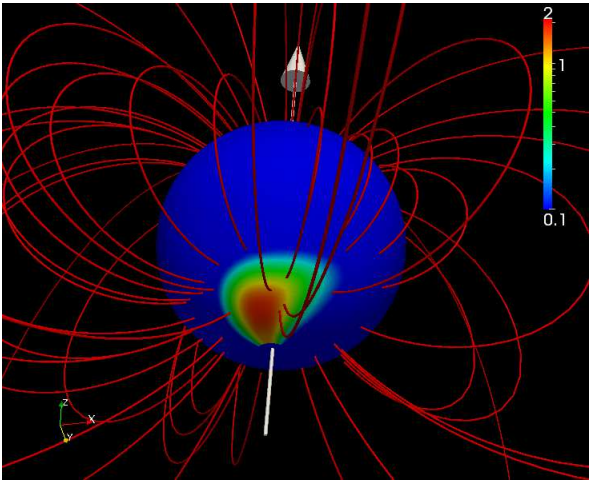}
  \caption{Particle number density distribution on the stellar surface, in
           log scale, in units of $10^{11}$~cm$^{-3}$ (see color table
           on the right of the panel) at the time $t=36$~hours. Selected
           magnetic field lines are overplotted in red. The white arrow
           marks the rotation axis of the star.}
  \label{acc_spot}
\end{figure}

The above results, therefore, show that the physical characteristics of
the accretion stream triggered by the flare closely recall those, largely
discussed in the literature, of streams driven by the accumulation of mass
at the disk truncation radius under the effect of the viscosity and pushed
out of the equatorial plane because of the growing pressure gradient there
(\citealt{2002ApJ...578..420R, Romanova2003ApJ, 2008A&A...478..155B,
2009A&A...508.1117Z}). Our simulation shows that the stream is relatively
cold (its temperature remains below 1 MK). After the disk material enters
the stream, its density slightly decreases and, close to the stellar
surface, increases again as a result of the gas compression by the dipolar
magnetic field. The stream velocity $u\rs{str}$ gradually increases toward
the star: it becomes supersonic already at a distance of $\approx 2\times
10^{11}$ cm from the disk, and $u\rs{str}$ approaches the free-fall speed
$u\rs{ff}$ close to the stellar surface. Figure~\ref{acc_spot} shows the
density distribution on the stellar surface at $t=36$~hours; the dense
spot on the surface is the region of impact of the stream. The spot covers
a small percentage of the stellar surface and the stream is inhomogeneous
with its mass density varying across the stream and being the largest
in the inner region, according to \cite{Romanova2004ApJ}.

\begin{figure}
  \centering
  \includegraphics[width=8.truecm]{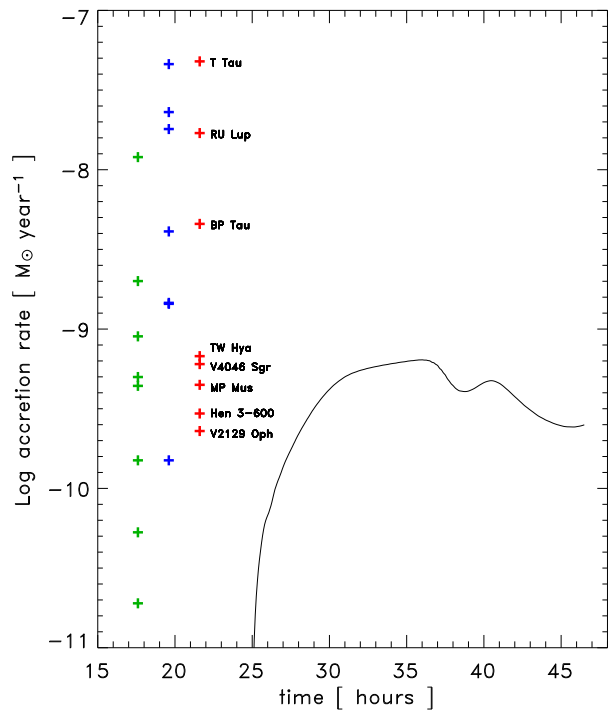}
  \caption{The solid line shows the mass accretion rate during the
           evolution of the funnel flow developed on the side of the
           disk opposite to the flaring loop. The crosses report the
           values of mass accretion rate derived from optical-near
           UV observations for a sample of low-mass stars and
           brown dwarfs (green; \citealt{2008ApJ...681..594H}),
           for a sample of solar-mass young accretors (blue;
           \citealt{2008ApJ...681..594H}) and for an X-ray-selected
           sample of CTTSs (red; \citealt{2011A&A...526A.104C}).}
  \label{fig3}
\end{figure}

Finally, we derived the mass accretion rate $\dot{M}$ due to the stream
from the side of the disk opposite to the post-flare loop and found
$\dot{M} \gsim 2.5\times 10^{-10}\, M\rs{\odot}$~yr$^{-1}$. Accretion
rates derived from optical-near UV continuum emission typically
range between $10^{-12}$ and $10^{-6}\, M\rs{\odot}$~yr$^{-1}$
with $\dot{M}$ varying in the same star even by a factor of 10 and
depending on the age and mass of the star (\citealt{2005ApJ...625..906M,
2005ApJ...626..498M, 2006A&A...452..245N}). In general, more evolved
low-mass young stars are characterized by lower accretion rates
(e.g. \citealt{2008ApJ...681..594H} and references therein). We
compared the rate $\dot{M}$ derived from our 3D simulation with
the rates derived from optical-UV observations and available in the
literature. In particular, we considered a sample of low-mass stars and
brown dwarfs observed with LRIS on Keck I and a sample of solar-mass
young accretors observed with HST STIS (both samples analyzed
by \citealt{2008ApJ...681..594H}), and an X-ray selected sample of
CTTSs observed with various optical telescopes (analyzed
by \citealt{2011A&A...526A.104C}). \cite{2008ApJ...681..594H}
found that the excess UV and optical emission arising in the Balmer
and Paschen continua yields mass accretion rates ranging between
$2\times 10^{-12}$ and $10^{-8}\,M\rs{\odot}$~yr$^{-1}$ in the case
of low-mass stars and brown dwarfs, and ranging between $2\times
10^{-10}$ and $5\times 10^{-8}\,M\rs{\odot}$~yr$^{-1}$ in the case
of solar-mass stars. \cite{2011A&A...526A.104C} calculated $\dot{M}$
for the stars of their sample by measuring H$\alpha$, H$\beta$,
H$\gamma$, He\,II (4686\,\AA), He\,I (5016\,\AA), He\,I (5876\,\AA)
O\,I (6300\,\AA), and He\,I (6678\,\AA) equivalent widths and found
that the mass accretion rates range between $2\times 10^{-10}$ and
$5\times 10^{-8}\,M\rs{\odot}$~yr$^{-1}$. Figure~\ref{fig3} shows the
rate $\dot{M}$ derived from our 3D simulation compared with the three
samples of young accreting stars. The accretion rate of our simulation
is in the range of the rates measured in low-mass stars and lower than
those of fast-accreting objects as BP Tau, RU Lup, or T Tau.

\section{Summary and conclusions}
\label{sec4}

We investigated the evolution of a bright flare occurring close to the
disk surrounding a magnetized CTTS and its effects on the stability of the
disk, through numerical MHD simulations. To our knowledge, the simulation
presented here represent the first attempt to model the 3D global
evolution of the star-disk system that simultaneously include stellar
gravity, viscosity of the disk, magnetic field, radiative cooling,
and magnetic-field-oriented thermal conduction (including the effects
of heat flux saturation). Our findings lead to several conclusions:

\begin{enumerate}

\item During its first stage of evolution, the flare gives rise to a
hot magnetic loop linking the inner part of the disk to the star; we
found that the length, maximum temperature, and peak X-ray luminosity
of the simulated flaring loop are similar to those derived from the
analysis of luminous X-ray flares observed in young low-mass stars
(\citealt{2005ApJS..160..469F}).

\item During the flare evolution, disk material evaporates in the outer
stellar atmosphere under the effect of the thermal conduction. A small
fraction of the evaporated disk material gradually fills the loop in
$\approx 10$ hours. Indeed most part of the evaporated disk material
is not efficiently confined by the magnetic field and channeled into
the loop but is rather ejected away from the central star in the outer
stellar corona, carrying away mass and angular momentum.

\item In the aftermath of the flare, the disk is strongly perturbed:
the injected heat pulse produces an overpressure in the disk at the
loop's footpoint that travels through the disk. When the overpressure
reaches the opposite side of the disk, a funnel flow starts to develop
there, flowing along the dipolar magnetic field lines and impacting onto
the stellar surface $\approx 1$ day after the injection of the heat
pulse. 

\item We found that the mass accretion rate $\dot{M}$ of the stream
triggered by the flare is in good agreement with those measured
in low-mass stars and brown dwarfs and in some solar-mass accretors
(e.g. \citealt{2008ApJ...681..594H, 2011A&A...526A.104C}). The stream
therefore accretes substantial mass (comparable with those observed) onto
the young star from the side of the disk opposite to the post-flare loop.

\end{enumerate}

We conclude therefore that the brightest flares detected in CTTSs
(e.g. \citealt{2005ApJS..160..469F}) can be a mechanism to trigger mass
accretion episodes onto protostars with accretion rates in the range of
those measured in low-mass stars and lower than those of fast-accreting
objects. On the other hand, it is worth mentioning that large flares do
not occur continuously in CTTSs and therefore cannot explain alone the
time-averaged accretion rates derived in young accretors: an average
of 1 flare per star per 650 ks ($\approx 7$ days) has been inferred by
analyzing the COUP observations (\citealt{2005ApJS..160..423W}) and
clusters much older than Orion (\citealt{2004ApJ...606..466W}). To
ascertain the contribution that flares may have to the observed
time-averaged accretion rates, it is necessary to determine the frequency
of those flares able to trigger accretion streams. In the future, an
exploration of the parameter space of our model is therefore needed to
determine to which extent the total energy released during the flare can
be reduced to still produce an accretion stream. This point is rather
important to ascertain if a storm of small-to-medium flares (very frequent
in CTTSs and often not even resolved in the lightcurves) occurring on
the accretion disk of young stars is able to trigger accretion streams
with high cadence, thus leading to a significant and persistent mass
accretion. This issue deserves further investigation in future studies.

The initial conditions adopted in our simulations are largely used in
the literature (e.g. \citealt{2002ApJ...578..420R} and subsequent works)
and are appropriate to describe the star-disk system in CTTSs. However,
as discussed in Sect.~\ref{evol_flare}, these conditions may result to
be idealized and simplified in some aspects. Many observations indicate
that the stellar atmospheres are permeated by magnetic fields with a
degree of complexity much higher than the magnetic dipole adopted here
(e.g. \citealt{2010RPPh...73l6901G} and references therein); the flare
may occur in the presence of existing accretion streams induced by the
coupling of the magnetic field to the disk below the corotation radius
(e.g. \citealt{2002ApJ...578..420R, Romanova2003ApJ, 2008A&A...478..155B,
2009A&A...508.1117Z}) or triggered by other flares. In particular,
the magnetic field in proximity of the heat release (due to magnetic
reconnection) is expected to be more complex than that modeled here,
thus confining more efficiently the hot plasma and enhancing the effects
of the flare energy deposition on the system. On the other hand, the
flaring loop and the flare-induced accretion stream could be strongly
modified or even disappear if the flare occurs in a flux tube initially
filled with cold infalling gas. However, \cite{2009pjc..book..179R}
showed that a flare involving dense infalling plasma would decay much
faster than generally observed, so that the flares we address here do
not occur in such a configuration. Nevertheless, the flaring loop and
the flare-induced accretion stream may be strongly perturbed and have
a more complex structure than that described here if the flare energy
release occurs in proximity of an existing accretion stream.

Following other works in the literature
(e.g. \citealt{2002ApJ...578..420R, Romanova2003ApJ, Romanova2004ApJ,
2008MNRAS.386..673K}), we assumed the turbulent diffusivity to be uniform
within the disk, setting $\alpha$ as a constant. However numerical
simulations indicate that the MRI stress would be substantial in the
disk midplane where the plasma $\beta$ is high, but would drop in outer
disk layers where $\beta$ is relatively low. As a result, the turbulent
diffusivity, scaled appropriately to the MRI, is expected to be smaller
in the outer disk layers than in the disk midplane. Since the $\alpha$
value adopted in our simulation (namely $\alpha=0.02$) lies in the lower
limit of the range of the parameter $\alpha$ (namely $0.01\leq \alpha
\leq 0.6$; \citealt{2003ARA&A..41..555B, 2008MNRAS.386..673K}), we expect
that the turbulent diffusivity would be larger in the disk midplane than
assumed in our simulation, thus influencing the mass accretion from the
disk onto the star. However, the time lapse covered by our simulation
($\approx 48$ hours) is $\approx 1/5$ of the star rotation period which
is much shorter than the time required by the turbulent diffusivity
(namely several star rotation periods; \citealt{2002ApJ...578..420R,
Romanova2003ApJ, 2008A&A...478..155B, 2009A&A...508.1117Z}) to be
effective and trigger magnetospheric funnel streams. In fact, throughout
the simulation, no stream develops in disk regions not influenced by
the flare although our model includes the disk viscosity.

As an additional check, we performed 2.5D simulations with the same
parameters of the 3D simulation but exploring the effect of $\alpha$
in the range between 0.02 and 0.1, and considering simulations either
with or without the disk viscosity. We found that the viscosity does not
significantly influence the results over the time laps covered by the 3D
simulation. We conclude therefore that a bright flare occurring close
to the circumstellar disk can trigger a substantial and long-lasting
accretion flow with physical characteristics similar to those caused
by the disk viscosity\footnote{Note that, as discussed above,
\cite{2009pjc..book..179R} has shown that the opposite is unlikely,
that is flares cannot be triggered in an accreting flux tube.}. This new
mechanism to drive mass accretion might be general. We suggest that the
flaring activity common to CTTSs may turn out to be an important factor
in the exchange of angular momentum and mass between the circumstellar
disk and the central protostar. This exchange is a rather hot and
controversial topic in astrophysics and our research contributes to
this issue, by exploring the role of the bright flares detected in CTTSs
(\citealt{2005ApJS..160..469F}) not considered up to now.

Our model predicts that the stream triggered by a flare impacts the
stellar surface with a delay of $\approx 1$ day since the peak X-ray
luminosity of the flare. This time delay depends on the physical
parameters of the model, namely the disk truncation radius, the mass
density and thickness of the disk, and the energy released by the heat
pulse. We expect therefore that the accretion shock generated by the
impact of the stream on the stellar surface might be visible with a
delay of the order of few days after the X-ray flare. Unfortunately, even
if this is the case, revealing both the X-ray flare and the delayed
accretion shock in real systems would be a rather difficult task because:
1) the star needs to be almost edge-on to allow both the flaring loop
and the accretion shock (located in different hemispheres of the star)
to be observable; 2) both the flaring loop and the accretion shock should
not be eclipsed by the star (or strongly absorbed by the circumstellar
material) during the observation. A more feasible task could
be to detect the effects of the X-ray irradiation of the disk by the
flare located just above it. For instance, a photoevaporation of disk
material by the X-ray emission from the flare is expected to be almost
instantaneous with the flare at variance with the case of a flare occurring
on the star which is assumed to be delayed given the distance of the
inner edge of the disk from the star. Searching for a tight correlation
between X-ray flares and variability in IR or optical band associated
with the photoevaporation of the disk could be an indication that the
flares were located close to the disk.

In addition to the results presented above, our simulation opens a number
of interesting issues. We have shown that the flaring loop linking the
disk with the star is a bright X-ray source, irradiating the disk from
above, at variance with the stellar corona which irradiates only the
innermost part of the disk. We suggest that this kind of sources may
strongly influence the chemical and physical evolution of the disk,
reaching parts of it that are effectively shielded from the stellar
coronal X-ray emission, with important consequences, for instance, on
the formation of planets. The turbulence and ionization state of plasma
resulting from the irradiance of the disk by the flaring loop may also
interact (because of the change of ionization and electrical conductivity)
in a nontrivial way with the MRI. Thus it is reasonable to suspect that
the effects of this X-ray radiation may influence the efficiency of MRI
and, therefore, the efficiency of angular momentum transport within the
disk. All these issues deserve further investigations.

\section*{Acknowledgments}
We thank an anonymous referee for constructive and helpful criticism.
We acknowledge stimulating discussions with C. Argiroffi, E. Flaccomio,
J. Kastner, A. Maggio, G. Micela, T. Montmerle, and S. Sciortino. PLUTO is
developed at the Turin Astronomical Observatory in collaboration with the
Department of General Physics of the Turin University. All computations
were performed on the IBM/Sp6 supercomputer at CINECA (Bologna, Italy),
and at the HPC facility (SCAN) of the INAF - Osservatorio Astronomico
di Palermo. This work was supported by the EU Marie Curie Transfer of
Knowledge program PHOENIX under contract No. MTKD-CT-2005-029768.

\bibliographystyle{mn2e}
\bibliography{biblio}

\label{lastpage}

\end{document}